\pdfoutput=1
\documentclass[twocolumn,10pt]{asme2e}
\usepackage{graphicx}
\usepackage{graphics}
\usepackage{color}
\usepackage{bm}
\usepackage{amsfonts,amssymb}

\confshortname{IMECE 2008}
\conffullname{ASME 2008 International Mechanical Engineering Congress and Exposition }
\confdate{31-06}
\confmonth{October-November}
\confyear{2008}
\confcity{Boston}
\confcountry{USA}

\papernum{IMECE2008-68026}

\title{Controlling Mixing Inside a Droplet by Time Dependent Rigid-body Rotation}

\author{Rodolphe Chabreyrie
    \affiliation{Mechanical Engineering Department\\
    Carnegie Mellon University\\
     Pittsburgh, Pennsylvania, 15213\\
    Email: rchabrey@andrew.cmu.edu
    }	
}

\author{Dmitri Vainchtein
    \affiliation{School of Physics\\
     Georgia Institute of Technology\\
     Atlanta, Georgia, 30332\\
     Space Research Institute\\
      Moscow, GSP-7, 117997, Russia\\
    }	
}
\author{Cristel Chandre
    \affiliation{Centre de Physique Th\'eorique$^0$\\
Luminy-case 907,\\
F-13288 Marseille cedex 09, France \\
 }
}

\author{Pushpendra Singh
    \affiliation{
    Mechanical Engineering Department\\
     New Jersey Institute of Technology\\
      Newark, New Jersey 07102\\
 }
}
\author{Nadine Aubry
    \affiliation{   Mechanical Engineering Department\\
    Carnegie Mellon University\\
     Pittsburgh, Pennsylvania, 15213\\
 }
}

\begin{document}

\maketitle

\begin{abstract}
{\it The use of microscopic discrete fluid volumes (i.e., droplets) as microreactors
for digital microfluidic applications often requires mixing enhancement and
control within droplets. In this work, we consider a translating spherical liquid
droplet to which we impose a time periodic rigid-body rotation which we model
using the superposition of a Hill vortex and an unsteady rigid body rotation.
This perturbation in the form of a rotation not only creates a three-dimensional
chaotic mixing region, which operates through the stretching and folding of
material lines, but also offers the possibility of controlling both the size and
the location of the mixing. Such a control is achieved by judiciously adjusting the three parameters that characterize the
rotation, i.e., the rotation amplitude, frequency and orientation of the rotation. As the size of the mixing region is
increased,
complete mixing within the drop is obtained.}
\end{abstract}
\footnotetext{UMR 6207 of the CNRS, Aix-Marseille and Sud Toulon-Var
Universities. Affiliated with the CNRS Research Federation FRUMAM (FR 2291).
CEA registered research laboratory LRC DSM-06-35.}


\section*{INTRODUCTION}
Although most microfluidics have been using fluid streams as the main means to carry fluids,
devices based on individual droplets have been proposed as well. In the latter,
also called ``digital microfluidic" systems, ``discrete'' fluid volumes
(droplets) rather than continuous streams are used, with the potential
to utilize individual droplets as microreactors \cite{Ismagilov:2003}.\\
Whether fluids are encapsulated within droplets or flow along channels, reactions can
occur efficiently only in presence of rapid mixing. Such mixing conditions are not easy
to fulfill due to the low Reynolds number of the flow involved, which prevents turbulence
from taking place. Stirring, in addition to molecular diffusion, is needed to stretch and fold
fluid elements, thus significantly increasing interfacial areas.  Whereas some strategies are
based on complex channel geometry for flows in microchannels, active methods
(via external forcing) (see, e.g. \cite{Oddy:2001,Bau:2001,Ouldelmoctar:2003,GlasgowAubry:2003,Glasgow:2004})
have also resulted in efficient mixing, especially at very low Reynolds numbers
\cite{Goullet:2006}. The combination of both geometry alteration and forcing has
been explored  as well \cite{Goullet:2006,Niu:2003,Bottausci:2004,Stremler:2004,Lee:2007}.\\
Chaotic advection inside a liquid droplet subjected to a forcing has been studied extensively
\cite{Baj,KroujilineandStone:1999,Lee:2000,WardandHomsy:2001,Grigoriev:2005,
Homsy:2007,VWG:2007} and obtained experimentally by means of oscillatory flows
\cite{WardandHomsy:2003,GSS:2006}. In this paper as in \cite{ChabreyriePRE08},
we concentrate on controlling both the size and the location of the mixing.
Indeed, we are interested in the ability to control the quantity of fluid mixed, from almost
zero mixing to complete mixing, as well as in the localization of the mixing.  On the one hand, researchers have mostly
concentrated so far on complete mixing as it is often desired for reactions to occur uniformly since variations in the mixing
could result in unacceptable variability in the properties and performance of the end-product.  On the other hand, incomplete
and
localized mixing - when well controlled - could be useful for the synthesis of inhomogeneous particles (see, e.g., Janus
particles) made of two (or more) components whose properties differ.  For instance, such particles could have a part of their
surface hydrophobic and some other parts hydrophilic.  Such a variation within the same particle could result in superior
particle properties useful in applications such as nanotechnology (self-assembly), biomedicine and advanced sensors.

In this article, we extend the work of
\cite{ChabreyriePRE08} that focuses on unsteady -- yet periodic -- forcing on a
translating droplet and its influence on the chaotic regions within the droplet.\\
The existence of chaotic behavior in three-dimensional bounded steady flows has
been shown (e.g. \cite{Baj,KroujilineandStone:1999,VVN:1996a,VNM:2006}). Our
work is distinct from the latter contributions through the addition of unsteadiness,
which is crucial to control the chaotic mixing behavior through resonance effects
\cite{Lima:1990,CFP2:1996,VWG:2007}.\\
The physical system under consideration in this work is described in the first section, and  numerical
results are exposed in the fourth section. We study the chaotic mixing
zone by monitoring  its location and its size. Specifically, the
creation and control of the chaotic mixing regions' location and size are
studied qualitatively via {\it Liouvillian sections}. A quantification of the
size as a function of the various parameters involved is also performed, thus
leading to mixing optimization in parameter space.  From this quantification of the size,
an optimization of the quantity of internal  mixing is performed.

\section*{PHYSICAL SYSTEM}
Let us assume a Newtonian liquid droplet of sperical shape suspended in an
incompressible Newtonian fluid whose motion consists of a translation and a slow
rigid body rotation (see \cite{KroujilineandStone:1999}). As in the
previous reference, we assume that the interfacial tension is sufficiently
large for the drop to remain spherical through its motion and that the Reynolds
number is very small compared to $1$. Thus, a reasonable approximation is to consider that
both the internal and external flows are Stokes flows. The boundary conditions at the droplet
surface can be derived from the continuity of velocity and tangential stress conditions.\\
The resulting internal flow is a superimposition of a steady base flow
(non-mixing flow) and an unsteady rigid-body rotation. Consider a Cartesian
coordinate system translating with the center-of-mass velocity of the droplet
with the orientation of the axes such that the unit vector ${\bm e}_{z}$ points in the
direction of the translation and the unit vector $\bm{e}_{x}$ lies in the
$\bm{\omega}-\bm{e}_{z}$ plane. We then obtain the following unsteady internal
flow:
\begin{eqnarray}
\label{veloT21}
u=\dot{x}&=& zx - \epsilon a_{\omega}(t) \sin\beta y,\\
\label{veloT22}
v=\dot{y}&=& zy + \epsilon a_{\omega}(t)\left(\sin\beta x - \cos\beta_x z\right),\\
\label{veloT23}
w=\dot{z}&=& 1-2x^2-2y^2-z^2 + \epsilon a_{\omega}(t)\cos\beta y,
\end{eqnarray}
In Eqs.~(\ref{veloT21}-\ref{veloT23}), all the lengths and velocities were made
dimensionless by using the droplet radius and the magnitude of the
translational velocity as the length and velocity scales, respectively.
The rotation is characterized by a maximum amplitude $\epsilon \ll
1$, a fixed orientation vector with angle $\beta$
and a form $a_{\omega}(t)$ given by
\begin{equation}
\label{cos}
a_{\omega}(t)=\frac{1}{2}\left(1+\cos \omega t \right),
\end{equation}

Note that the equations above, together with their derivation, are identical to those of
\cite{KroujilineandStone:1999}, except that the vorticity vector is now time dependent.  The time dependency is introduced
either
via the external boundary conditions or through a time dependent body force in the momentum equation.

\begin{figure}[t]
\begin{center}
\includegraphics*[width=7cm]{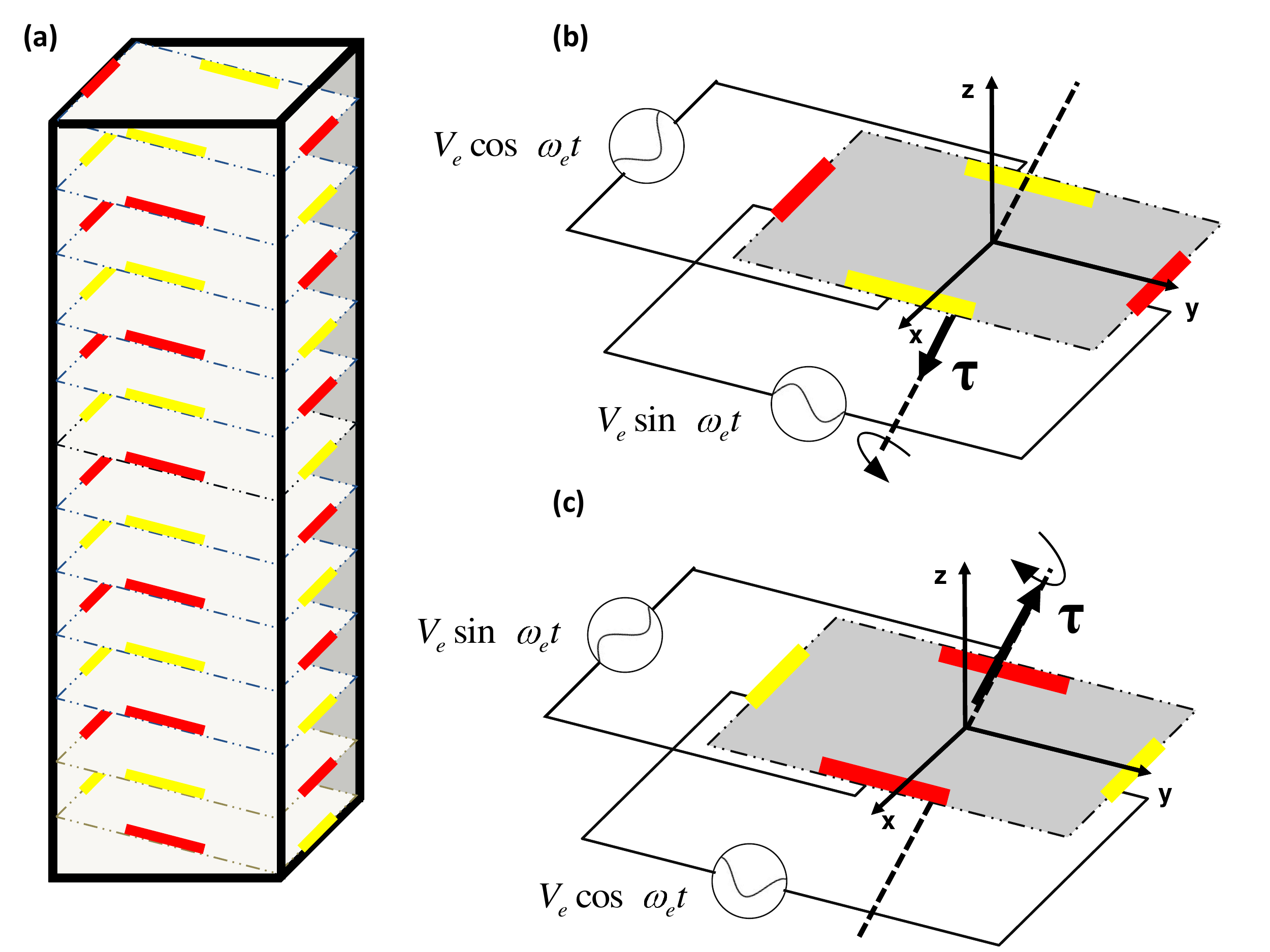}
\end{center}
\caption{SKETCH OF A POSSIBLE EXPERIMENTAL APPARATUS SHOWING (a) A SQUARE CHANNEL OR COLUMN WITH A SERIES OF OUT-OF-PHASE
ELECTRODES EMBEDDED IN THE FOUR WALLS. A TORQUE ACTS ON THE DROP LOCATED AT THE ORIGIN OF THE FRAME OF REFERENCE IN EACH PLANE.
SUCH A TORQUE IS IN THE DIRECTION PERPENDICULAR TO EACH PLANE, WHICH LIES AT AN ANGLE WITH THE STREAMWISE DIRECTION OF THE
CHANNEL (Z).  THE VOLTAGES ENERGIZING THE ELECTRODES ALTERNATE BETWEEN THE VARIOUS PLANES (AS SHOWN IN THE FIGURE) SO THAT THE
TORQUE SWITCHES BETWEEN POSITIVE AND NEGATIVE VALUES AND THE DROPLET UNDERGOES ACCELERATING AND DECELERATING ROTATION AS IT
TRANSLATES.
 (b) ACCELERATING ROTATION (POSITIVE TORQUE $\tau>0$). (C) DECELERATING ROTATION (NEGATIVE TORQUE $\tau<0$).
 $V_e$ AND $\omega_e$ ARE THE VOLTAGE AND FREQUENCY APPLIED TO THE ELECTRODES. THE DASH LINE REPRESENTS THE AXIS OF ROTATION.}
\label{figure0}
\end{figure}

In practice, this could be realized for instance
by creating a time-dependent swirl motion in the external flow or by
applying an electric field that exerts a torque on the drop (see,
e.g., work on traveling wave dielectrophoresis \cite{AubrySingh:2006} or on electrorotation \cite{Arn:1988}). A possible design
based on the latter phenomenon is proposed in
Fig.~\ref{figure0}.  Electrorotation stands for the spinning of an electrically polarized particle (or droplet in our case)
while
the latter is subjected to a ``rotating'' electric field, that is an ac electric field generated by voltages which are out of
phase of one another.
The proposed device would thus consist of a square column/channel with electrodes embedded within its four walls creating a
rotating electric field due to the phase difference between the voltages of adjacent electrodes.  It is known that such a
four-pole electrode setting would generate a controlled spinning motion, or rigid body rotation, of a drop trapped in the middle
of the channel.  The drop steady translating motion, on another hand, could simply be produced by a constant pressure gradient
along the channel or simply buoyancy in the case of a vertical column. In order to vary the
orientation of the axis of rotation compared to the direction of translation, we propose a series of electrodes lying within
inclined planes along the length of the channel (see Fig.~\ref{figure0}).  The angle of the inclined planes with the vertical
could be adjusted. The periodicity in the angular velocity is then obtained by imposing a phase difference to the voltages
applied to the electrodes of adjacent planes, thus generating acceleration and deceleration phases in the imposed rigid body
rotation.

\section*{NON-MIXING CASE}
The non-mixing (unperturbed) case or base flow, i.e. $\epsilon =0$, is characterized
by streamlines that are joint lines of constant streamfunction $\psi$ and azimuthal angle $\phi$, denoted by
$\Gamma_{\psi,\phi}$.
\begin{equation}
\psi = \frac{1}{2} \left(x^2+y^2 \right)\left(1-x^2-y^2-z^2\right), \quad \phi =\arctan \left(y/x\right),
\nonumber
\end{equation}
where $\psi \in \left[0,1/8\right]$ and $\phi \in [0,2\pi[$.
Besides the heteroclinic orbits connecting the  two hyperbolic fixed
points located at the poles of the sphere, all other streamlines are closed curves that converge
toward a circle of degenerate elliptic fixed point $(x^2+y^2=1/2, z=0)$ as
$\psi$ is increased toward the value $1/8$.
The frequency of motion on the
streamline $\Gamma_{\psi,\phi}$ is independent of $\phi$ and given by
\begin{equation}
\label{Per}
\frac{2\pi}{\Omega(\psi)}  =\frac{2\sqrt{2}}{\sqrt{1+\gamma}}
K\left( \sqrt{\frac{2\gamma}{1+\gamma}}\right),
\end{equation}
with $\gamma(\psi)=\sqrt{1-8\psi}$ and $K$ is the complete elliptic function
of the first kind.
The trajectories of the system (\ref{veloT21}-\ref{veloT23}) are periodic orbits (with period $2\pi/\Omega(\psi)$)
which come in families (labeled as $\psi$). No chaotic mixing occurs since there is no exponential divergence in
the bulk of the droplet. However, due to the difference in frequency a very small
mixing occurs in $\psi$ (but not in $\phi$ due to the degeneracy).
\section*{MIXING CASE}
In this work we follow the approach of  \cite{ChabreyriePRE08} that
enables the generation of a three-dimensional chaotic mixing region,
for which we are able to control both the location and the size. The method consists in
bringing a specific family of the unperturbed tori $ \left\{
\Gamma_{\psi_{n}}\right\}_{n~\in \mathbb{N}}$  (composed by a family of periodic orbits along the azimuthal angle) into
resonance with the periodic perturbation $a_{\omega}(t)$ by
selecting the frequency $\omega$ so that it satisfies the resonance condition:
\begin{equation}
\label{resonance}
n\Omega\left(\psi_{n}\right)-\omega=0,~~\mbox{for}~~n \in \mathbb{N}.
\end{equation}
The control is realized by adjusting the three parameters that
characterize the periodic rigid body rotation, specifically its the maximum
amplitude $\epsilon$, its frequency $\omega$ and the orientation of the axis of rotation $\beta$
(see Eqs.~\ref{veloT21}-\ref{veloT23}). The amplitude satisfies $0\leq\epsilon<<1 $, where
the lower and upper limits correspond to the absence of mixing and
complete mixing, respectively.
 \begin{figure}[t]
\begin{center}
\includegraphics*[width=7cm]{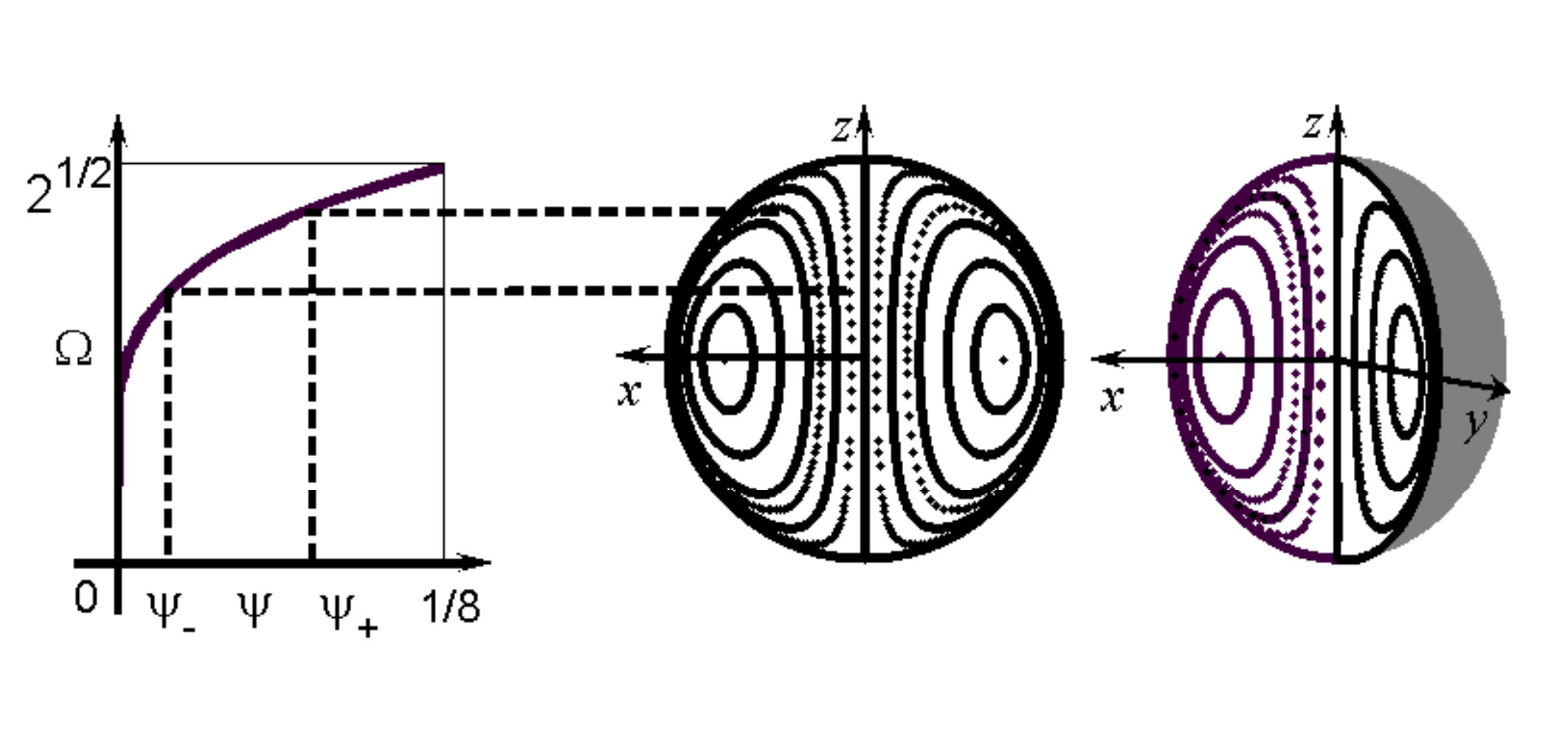}
\end{center}
\caption{STREAMLINES IN A CROSS-SECTION OF THE DROPLET  (WITHOUT ROTATION) AND THEIR
FREQUENCIES  $\Omega\left(\psi\right)$ AS GIVEN BY
Eq.~(\ref{Per}). REPRINTED FROM \cite{chabreyrieMRC08} WITH PERMISSION FROM ELSEVIER.}
\label{figure1}
\end{figure}
\\
\\
\\
\section*{NUMERICAL RESULTS\label{results}}
The numerical results presented below have been obtained by using a standard explicit fourth order Runge-Kutta scheme
\cite{RK4}.\subsection*{Controlling the location of the chaotic mixing region\label{location}}

\begin{figure}[t]
\begin{center}
\includegraphics*[width=7cm]{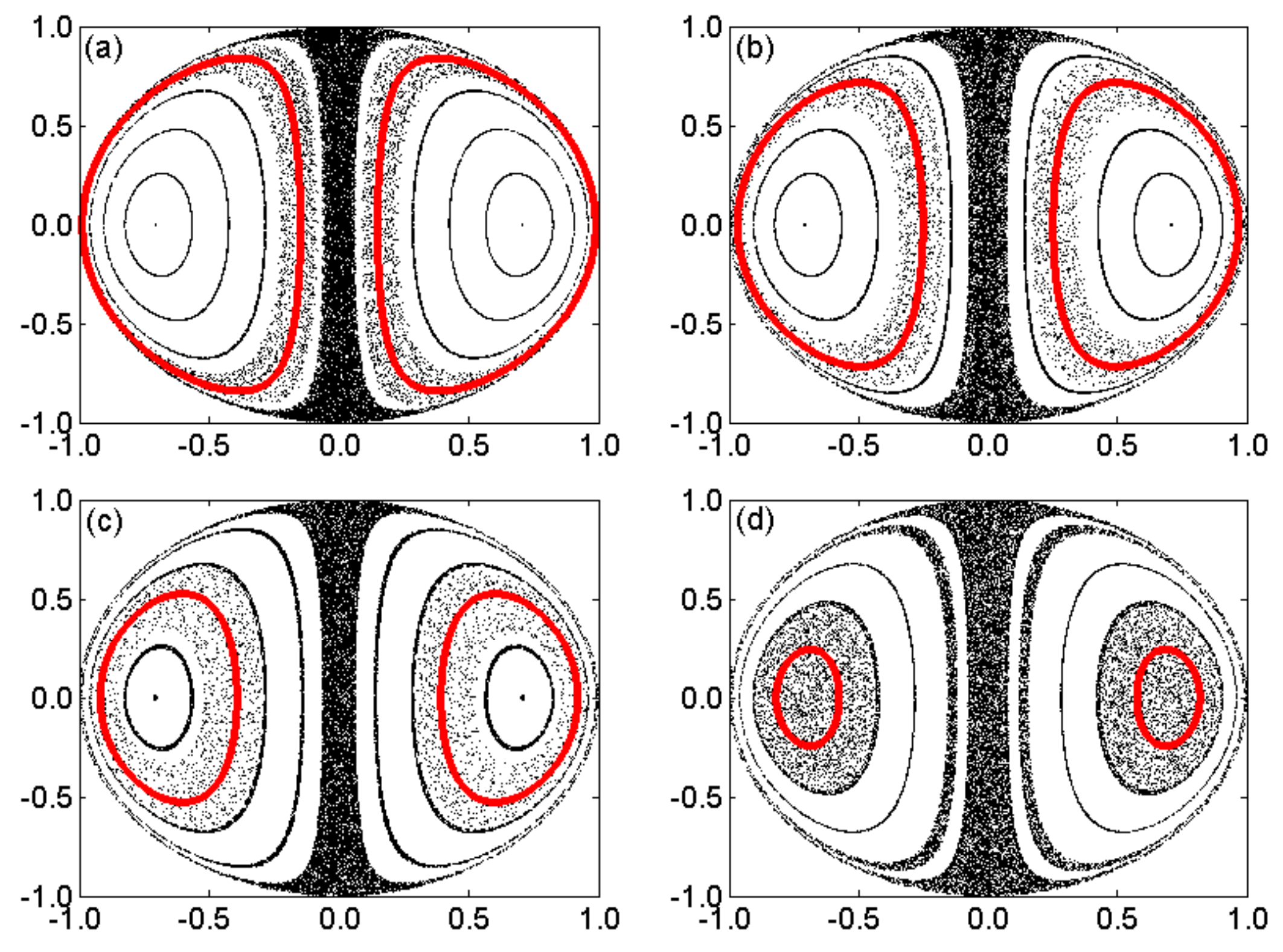}
\end{center}
\caption{LIOUVILLIAN SECTIONS FOR THE FREQUENCIES $\omega= 0.95, 1.1, 1.25, 1.40$ (a-d),
THE AMPLITUDE $\epsilon = 0.05$ AND FOR ORIENTATION $\alpha = \pi/4$. THE (RED)
LINE INSIDE THE $CMR _1$ IS THE TORUS $ \Gamma_{\Omega^{-1}(\omega)}$. REPRINTED FROM \cite{chabreyrieMRC08} WITH PERMISSION FROM ELSEVIER.}
\label{figure2}
\end{figure}

Figures~\ref{figure2},~\ref{figure3}  display the {\it
Liouvillian sections} of the mixing system, which consist of two-dimensional projections
of time-periodic three-dimensional flows by a combination of a stroboscopic map
and a plane section (here, the $y=0$ plane). In other words, the points on the Liouivillian sections are
the intersections of the trajectories with the plane $y=0$ at every period $2\pi/\omega$.\\
The perturbation $a_{\omega}(t)$ generates two
non-negligible three-dimensional chaotic mixing regions (see Fig.~\ref{figure2}):
one around the torus  having the frequency $\omega$ labeled  $CMR_1$ and another one around the
pole-to-pole axis and near the droplet surface labeled $CMR_{n>1}$.
In Fig.~\ref{figure2},  we clearly see that  the location of $CMR_1$  varies with the value
of $\omega$ according to Eq.~(\ref{resonance}). It is important to note that $CMR_{n>1}$ stay anchored
around the pole-to-pole axis  and to  the surface due to the flat part  of $\Omega(\psi)$ close
to $\psi=0$. This is due to the flat part of $\Omega(\psi)$ (see Fig.~\ref{figure1}). \\
For small values of $\omega$, all resonances are located near the pole-to-pole
heteroclinic connections (at $\psi=0$, near the $z$ axis and near the surface
of the droplet, see Fig.~\ref{figure2}). For larger $\omega$ values, $CMR_1$
separates from the chaotic mixing region close to the pole-to-pole axis and
penetrates deeper into the droplet. In the interval $0<\omega<
\sqrt{2}$, $CMR_1$ is the largest chaotic region, followed by $CMR_{n>1}$ .
As $\omega$ is increased further, $CMR_1$ moves toward the location of the
elliptic fixed point of the unperturbed system by following the location of the
resonant tori with the frequency $\omega$.

\subsection*{Controlling and optimizing the size of the chaotic mixing region\label{size}}

\begin{figure}[t]
\begin{center}
\includegraphics*[width=7cm]{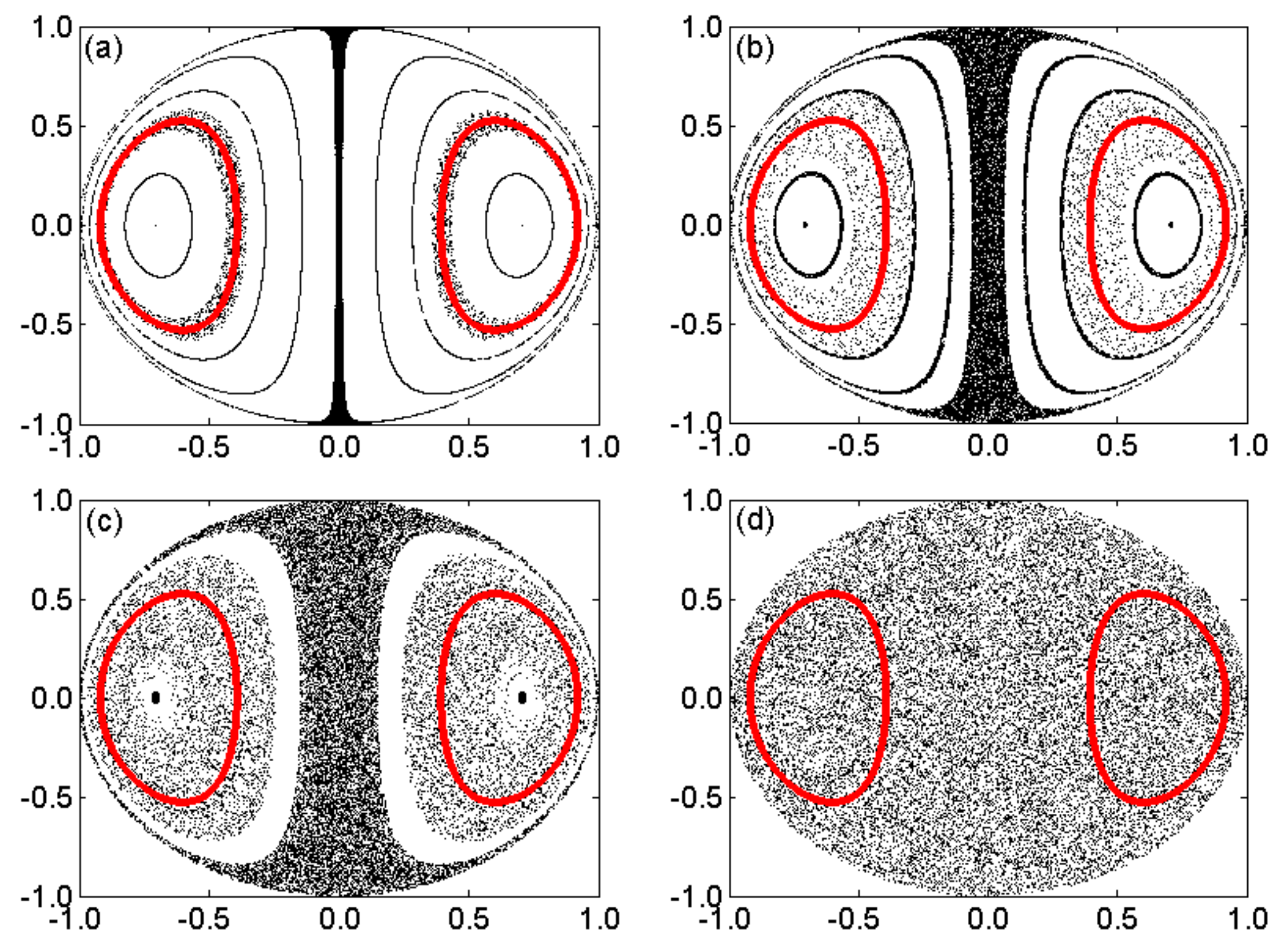}
\end{center}
\caption{LIOUVILLIAN SECTIONS FOR
THE  FREQUENCIES $\omega = 1.25 $, THE AMPLITUDE $\epsilon = 0.01, 0.05, 0.10, 0.20$
(a-d) AND FOR THE  ORIENTATION $\alpha = \pi/4$. THE  (RED) LINE INSIDE THE $CMR_1$
IS THE TORUS  $\Gamma_{\Omega^{-1}(\omega)}$. REPRINTED FROM \cite{chabreyrieMRC08} WITH PERMISSION FROM ELSEVIER.}
\label{figure3}
\end{figure}

In this section, we quantify the size of the  main chaotic mixing region,
i.e., $CMR_1$ as the three parameters $\epsilon$, $\omega$ and
$\beta$ vary. This quantification  is performed by computing the the maximum variation
$\Delta \psi$ of the stream function $\psi$ for one trajectory inside the $CMR_1$
Whereas the frequency $\omega$ of the rigid body rotation is mostly responsible
for the location of $CMR_1$ (while $CMR_{n>1}$ is always anchored around the
heteroclinic orbits), it is the amplitude $\epsilon$ and the orientation $\beta$
that mostly determine the size of the $CMR_1$.
Figure~\ref{figure3} shows qualitatively the size of the chaotic mixing regions
created by the $n=1$ and, $n>1$ resonances, i.e., $CMR_1$ and
$CMR_{n>1}$. In this figure, we see that the size of both the $CMR_1$ and $CMR_{n>1}$ increases
with the amplitude of the perturbation. It is also interesting to note that around
$\epsilon=\epsilon_{max}\approx 0.20$,  the $CMR_1$ and $CMR_{n>1}$
join together to cover the entire droplet volume. At that point, complete
chaotic mixing is obtained. The size of $CMR_1$  as a
function of the frequency $\omega$ is depicted in Fig.~\ref{figure5} (lower
panel). From this figure  it is clear that for each value of $\epsilon$ the size reaches a
maximum for a certain value $\omega_{max}$ of the forcing frequency.
\begin{figure}[t]
\begin{center}
\includegraphics*[width=7cm]{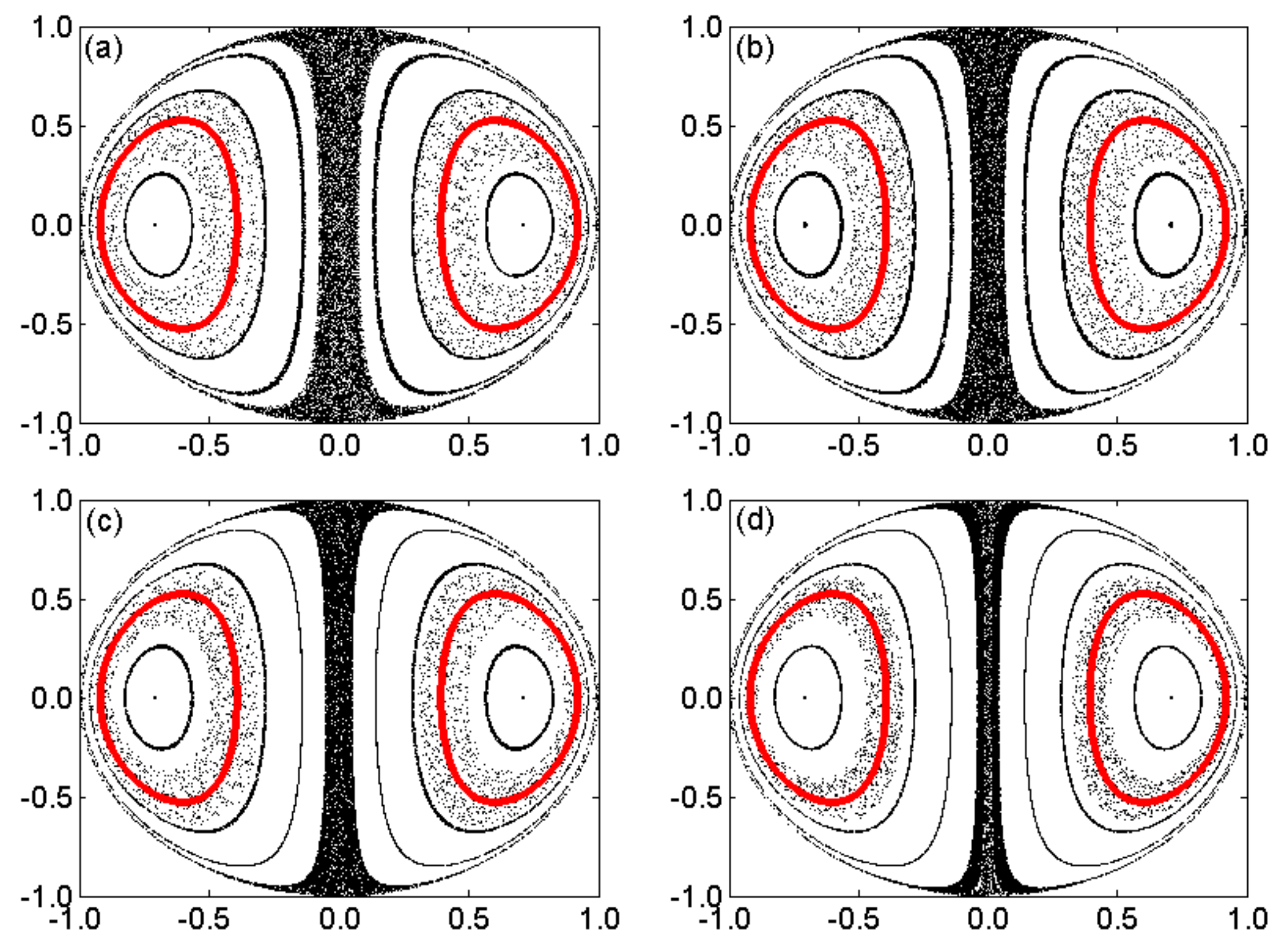}
\end{center}
\caption{LIOUVILLIAN SECTIONS FOR THE FREQUENCY  $\omega = 1.25 $, THE  AMPLITUDE $\epsilon =
0.05$ AND THE ORIENTATIONS  $\alpha = \pi/8, \pi/4, 29\pi/80, 7\pi/16$ (a-d).
THE (RED) LINE INSIDE  THE $CMR _1$ IS THE TORUS  $\Gamma_{\Omega^{-1}(\omega)}$.}
\label{figure4}
\end{figure}

The effects of the parameter $\beta$ on the size of $CMR_1$ present two distinguishable behaviors
one where the size is weakly dependent and another where the size is strongly dependent.
Such behaviors are easily observed on the lower panel of Fig.~\ref{figure5}
where  the size of the $CMR_1$  as a function of $\beta$ is displayed.
On one hand, when $\beta$ is well inside  the two limit values,  $0$ and $\pi/2$,
the size is practically constant.  On the other hand, when $\beta$ gets closer to the limits,
the size decreases significantly.
Such observation are qualitatively confirmed in the Liouvillian section in Fig. \ref{figure4}.
In the first two panels,  we see a very small change in size.  However, in the last two panels the
size decreases drastically. Having  such a relation between the size of the $CMR_1$
and the orientation of the rotation $\beta$ can be handy in practice to control
the size of mixing. In some applications, where the experimenters are faced with
the dilemma of precisely controlling the size of mixing  with an imprecise fluctuation
of $\beta$, setting  $\beta$ well inside the limit value and controlling the size through
 the parameter $\epsilon$ could be the solution. Then, more restrictive, applications
where increasing the amplitude of rotation   may not be possible (e.g. biomedical ones where
biological particles need to be handled with care), fixing $\epsilon$  and tuning the size of
the mixing by varying $\beta$ around the limit value $\beta=\pi/2$  could be the solution.
One should notice that at the critical orientation  $\beta=\pi/2$ no three-dimensional chaotic
mixing occurs since $\psi$  is conserved.
\begin{figure}[t]
\begin{center}
\includegraphics*[width=7cm]{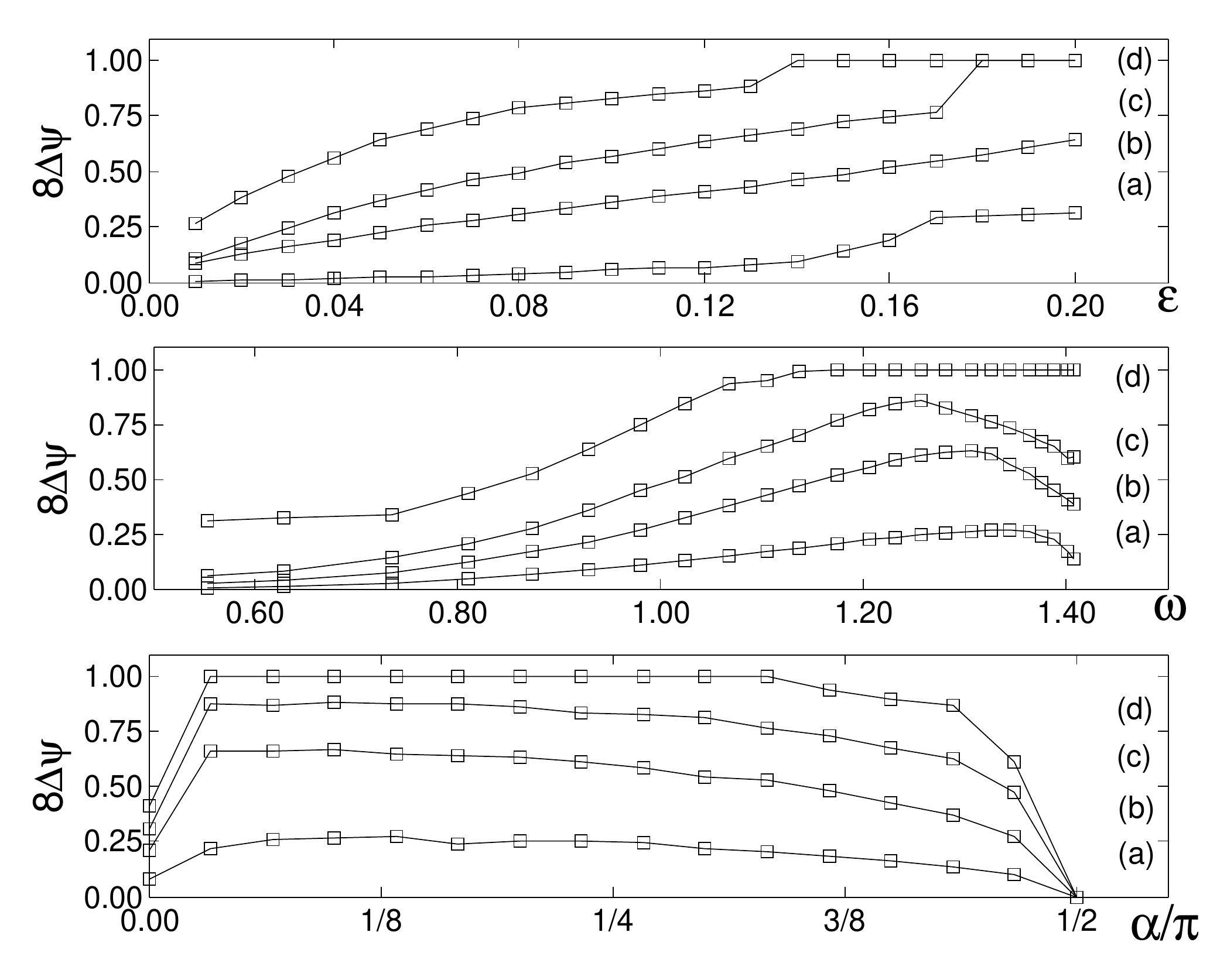}
\end{center}
\caption{UPPER PANEL: NORMALIZED SIZE $\Delta\psi$ {\it VS.} FREQUENCY $\omega$ FOR
AMPLITUDES $\epsilon = 0.01, 0.05, 0.10, 0.20$ (a-d) WITH  ORIENTATION $\beta =
\pi/4$; MIDDLE PANEL: NORMALIZED SIZE $\Delta\psi$ {\it VS.}
AMPLITUDE  $\epsilon$ FOR  FREQUENCIES $\omega = 0.55, 0.93, 1.28, 1.41$ (a-d) WITH
THE ORIENTATION $\beta = \pi/4$; LOWER PANEL: NORMALIZED SIZE $\Delta\psi$ {\it VS.}
ORIENTATION $\beta$ FOR  AMPLITUDES $\epsilon = 0.01, 0.05, 0.10, 0.20$ (a-d) WITH
FREQUENCY $\omega = 1.25$. TOP AND MIDDLE PANELS REPRINTED FROM \cite{ChabreyriePRE08} WITH PERMISSION FROM THE AMERICAN PHYSICAL SOCIETY.}
\label{figure5}
\end{figure}
\section*{CONCLUSIONS}
In this work we have shown that chaotic mixing within a translating droplet can be
obtained by adding a perturbation in the form of an oscillatory rigid body
rotation. The frequency of the latter was selected in order to create
resonances with the natural frequencies of the system, namely the frequencies
of the various tori embedded within the droplet. A particularly interesting
feature of the perturbed system lies in the fact that both the size and the
location of the mixing region can be varied. This was achieved by adjusting the
control parameters of the perturbation, specifically the frequency, amplitude
and orientation of the rigid body rotation. While the frequency is used to
target a particular location, both the amplitude and the orientation influence
the size.  The latter property introduces additional flexibility in the system,
particularly for applications which allow for the variation of one of these two
parameters only.
\bibliographystyle{asmems4}
\begin{acknowledgment}
his article is based upon work partially supported by the NSF (grants
CTS-0626070 (N.A.), CTS-0626123 (P.S.) and 0400370 (D.V.)). D.V. is grateful to
the RBRF (grant 06-01-00117) and to the Donors of the ACS Petroleum Research
Fund. C.C. acknowledges support from Euratom-CEA (contract EUR~344-88-1~FUA~F)
and CNRS (PICS program).
\end{acknowledgment}
\bibliography{biblio}

\end{document}